\documentclass[prb,twocolumn,showpacs,floatfix,amsfonts]{revtex4}
\usepackage{graphicx,graphics,color,epsfig}
\usepackage{bm}
\usepackage{amsmath}
\usepackage{amssymb}
\begin{document}
\preprint{}
\title{Fourier-Transformed Local Density of States and Tunneling into a
$D$-Wave Superconductor with Bosonic Modes}
\author{Jian-Xin Zhu and A. V. Balatsky}
\affiliation{Theoretical Division, MS B262, Los Alamos National
Laboratory, Los Alamos, New Mexico 87545}
\author{T. P. Devereaux}
\affiliation{Department of Physics, University of Waterloo, Ontario,
Canada N2L 3GI}
\author{Qimiao Si}
\affiliation{Department of Physics \& Astronomy, Rice University,
Houston, TX 77005}
\author{J. Lee, K. McElroy, and J. C. Davis}
\affiliation{LASSP, Department of Physics, Cornell University,
Ithaca, NY 14850}

\date{May 28, 2005}

\begin{abstract}
We analyze the effects of the electronic coupling to bosonic modes
in a d-wave superconductor. The role of the scattering due to boson
on  the momentum transfer between  electronic states in the
Brilloine zone is addressed. We consider specific examples of
$B_{1g}$ phonon, breathing mode phonon and spin resonance at
$(\pi,\pi)$.  The Fourier spectrum of the energy derivative local
density of states (LDOS) is calculated. To properly calibrate the
effects of different modes we fix the quasipartilce renormalization
at specific momentum points. It is found that the $B_{1g}$ mode
with highly anisotropic momentum-dependent coupling matrix element
gives rise to well definded features in the Fourier spectrum, at the
energy of mode plus gap, with a momentum transfer along the Cu-O
bond direction of cuprates. This result is  in  a striking contrast
to the cases of the coupling to other modes and also to the case of
no mode coupling. The origin of this difference is explored in
detail. A comparison with the recent STM experiments is briefly
discussed.

\end{abstract}
 \maketitle

\section{Introduction}
Determining the nature of single particle excitations is of
fundmental importance in our understanding of the superconductivity
in high-$T_c$ cuprates.   To address this issue, a number of
spectroscopies have been extensively used, including
the angle resolved photoemission spectroscopy (ARPES) and tunneling.
The salient features observed in ARPES include: (a) Near the
$(\pi,0)$ $(M)$ point in the Brillouin zone, the spectral function
in the superconducting state shows an anomalous line shape, the
so-called ``peak-dip-hump''
structure;~\cite{Dessau91,Ding96,Campuzano99} (b) Near the $d$-wave
node of the superconducting gap, the dispersion shows a
characteristic ``kink'' near 50-70
meV.~\cite{Bogdanov00,Kaminski01,Lanzara01,Johnson01,Zhou03} Recent
ARPES experiments with improved
resolution~\cite{Kim03,Gromko03,Sato03,Cuk04} have revealed another
``kink'' in dispersion of the antinodal electronic states, near the
$M$ point. An unusual spectral dip-hump features similar to the
ARPES spectrum have also been observed in the tunneling
data.~\cite{Huang89,Renner95,Renner96,DeWilde98,
Mandrus91,Yurgens99,Zasadzinski00,Zasadzinski01} All these features
were suggested to indicate that the electron self-energy
renormalization could be  due to the electronic coupling to a
bosonic mode. Two main scenarios have been presented to explain the
experimental data. On  one hand, the antinodal renormalization is
found to be strongly enhanced below
$T_c$.~\cite{Kim03,Gromko03,Sato03} Such a strong temperature
dependence and the dominance of the coupling strength near the $M$
point can be thought of as  evidence for the  coupling of electrons
to the 41 meV spin resonance mode (of electronic
origin).~\cite{Dahm96,Shen97,Norman97,Norman98,Abanov99,Eschrig00,Norman01,
Manske01,Abanov03} As seen by inelastic neutron scattering
experiments in most of the
cuprates,~\cite{Rossat-Mignod91,Mook93,Fong95,Bourges95,Bourges96,Fong99,Dai96,
Fong97,Bourges97,Dai99,He01,He02,Bourges98,Fong00,Bourges00,Dai01}
the spin mode intensity substantially turns on below $T_c$ (even
though some intensity might be present in a normal state) and has a
well-defined momentum of $(\pi,\pi)$. This scenario of the
electronic coupling to spin resonance mode has also been used to
explain the tunneling spectra in planar tunnel junctions. On the
other hand, it has been suggested that a significant electronic
coupling to the half-breathing in-plane Cu-O bond stretching phonon
or to the out-of-plane out-of-phase O buckling $B_{1g}$ phonon, with
an energy of approximately 70 and 35 meV, respectively, might be
responsible for the dispersion anomalies at the
nodal~\cite{Lanzara01} and antinodal directions,~\cite{Cuk04}
respectively. These two phonon modes have shown strong lineshape
renormalizations with doping and temperature in Raman and neutron
measurements.~\cite{Friedl90,Pyka93,
Reznik95,Chaplot95,McQueeney99,Chung03,Sugai03,Opel99}
The advantage of this scenario is   that it could naturally explain
the band renormalization effect in materials where no spin resonance
mode has been detected, in the normal state, and in the deeply
overdoped region where the spin mode is neither expected nor
observed. To be consistent with ARPES data,
this scenario requires
the electron-phonon interaction to be highly
anisotropic~\cite{Sandvik04,Devereaux04} and its impact on the
electrons to be strongly enhanced in the superconducting state.


The nature of the involved bosonic modes, being phononic or
electronic and their role in the mechanism of superconductivity
remains controversial. Detailed energy and momentum spectoscopy of
the relevant bosonic modes might be very helpful in understanding
the mechanism of superconductivity in high-$T_c$ cuprates. One of
the direct spectroscopies that allows energy determination is
Inelastic Electron Tunneling Spectroscopy (IETS). It  is a
well-established and powerful tool that allows the measurement of
the  characteristic energies of extended modes. Examples of
applications of this technique, among many, include measurements of
molecular stretching and vibrational modes in metal-insulator-metal
tunnel junctions;~\cite{Jakievic66,Hansma82}  observation of the
collective magnetic resonance in tunneling in the superconducting
state of high-$T_c$ materials~\cite{Zasadzinski01,Norman97} and
observation of the tunneling features at energies that correspond to
the phonon peaks, as seen in planar tunneling into
superconductors.~\cite{Scalapino69}

The IETS directly measures excitation energies. When electrons
scatter off a collective mode, a contribution to the electron self
energy occurs above a corresponding threshold value of the frequency
determined by the mode frequency. Thus, in a tunneling experiment,
for bias voltages exceeding the threshold, electrons can excite the
mode. This additional scattering channel leads to a step in the
density of states (DOS) and in the tunneling conductance. Low
temperatures are required to avoid thermal smearing of the step in
the conductance. The crucial quantity that reveals inelastic peaks
is a second derivative of tunneling conductance with respect to bias
voltage: $\frac{d^{2}I}{dV^{2}}(eV)$. Peaks in this quantity are
shown to be connected to the energies of the modes, e.g  peaks in
phonon Density of States.~\cite{Scalapino69}

In high-$T_c$ materials the energies of a  number of collective
modes, like phonon and spin modes are close. For example the
$B_{1g}$ phonon mode has typical energy of 36-40 meV  and spin
resonance mode has an energy that ranges between 35-40 meV depending
on doping. Hence discrimination between different modes based on
only the energy of the observed mode is a challenge. Aside from
energy resolved  features,  it would be useful to come up with the
measurement that would allow one to measure typical momenta involved
in the electron scattering in cuprates. The ARPES is one
such spectroscopy.

Scanning Tunneling Spectroscopy (STM) is another technique that
would allow one to resolve the momentum transfer between different
electronic states. Recently, the Fourier-transformed scanning
tunneling microscopy (FT-STM)~\cite{Hoffman02b,McElroy03} was
introduced to  map out the Fermi surface and the momentum
dependence of the $d$-wave energy gap. The results are in good
agreement with the ARPES data. While the ARPES provides the
information about the energy dispersion of single-particle
electronic states, the FT-STM probes the scattering  processes
between  states with different in plane momenta.

In this paper, we show how another technique, the Fourier Transform
Inelastic Electron Tunneling Spectroscopy (FT-IETS), would allow
simultaneous momentum and energy resolutions of the tunneling electrons.
As such, the technique might be useful to address the role
of different modes in cuprates. Motivated by the
progress of elastic FT-STM  technique, earlier few of us (JXZ, QS,
AVB) have suggested an FT-IETS technique, Ref.~\onlinecite{Zhu04},
in  a specific model that describes the Fourier transformed local
density of states (LDOS) in a $d$-wave superconductor with the
electronic coupling to the $(\pi,\pi)$ spin resonance mode. Here we
go beyond the previous analysis and address the question of what
consequences the electron self-energy renormalization will have on
the tunneling characteristics for a number of collective modes
broadly considered in the literature: the $B_{1g}$ and breathing
phonon modes, as well as the $(\pi,\pi)$ spin resonance mode. The central
quantity we will focus on in our analysis is   energy derivative of the
FT LDOS, $\rho^{\prime}(\mathbf{q},E)$. This quantity corresponds to
$\frac{d^{2}I}{dV^{2}}(\mathbf{q},eV)$ measured by the FT-STM, where
$I$ is the local tunneling current and $V$ the voltage bias. Here we
are not concerned with the mechanism of the superconductivity in the
cuprates. Instead we assume from the outset a $d$-wave channel
effective pairing interaction and study the additional effects due
to the electronic coupling to various bosonic modes, including  the
$B_{1g}$ and breathing phonon modes, and the $(\pi,\pi)$ spin
resonance mode. A comparison of the calculated momentum transfer
structure with the FT-STM measurement may shed new light on which
type of bosonic mode the electron excitations are coupled to
strongly.

The rest of the paper is outlined as follows: In
Sec.~\ref{SEC:Model}, we develop a theoretical model in which the
electrons are coupled to bosonic modes. We consider three types of
modes: the $B_{1g}$ and breathing phonon modes with possible
coupling matrix elements, and the $(\pi,\pi)$ spin resonance mode.
Additional weak disorder is used as a marker so that
the momentum transfer can be investigated.  In
Sec.~\ref{SEC:Results}, the numerical results for the FT spectrum of
the energy derivative LDOS are presented.
Sec.~\ref{SEC:Conclusion} contains some concluding remarks.

\section{Theoretical model}
\label{SEC:Model}

We start with a BCS-type model to describe two-dimensional electrons
with a $d$-wave pairing symmetry, which is relevant to
high-temperature cuprates:
\begin{equation}
\mathcal{H}_{BCS}=\sum_{\mathbf{k},\sigma} \xi_{\mathbf{k}}
c_{\mathbf{k}\sigma}^{\dagger} c_{\mathbf{k}\sigma}
+\sum_{\mathbf{k}}(\Delta_{\mathbf{k}}
c_{\mathbf{k}\uparrow}^{\dagger}c_{-\mathbf{k}\downarrow}^{\dagger}
+\Delta_{\mathbf{k}}^{*}
c_{-\mathbf{k}\downarrow}c_{\mathbf{k}\uparrow})\;,
\end{equation}
where $c_{\mathbf{k}\sigma}^{\dagger}$ ($c_{\mathbf{k}\sigma}$)
creates (annihilates) a conduction electron of spin $\sigma$ and
wavevector $\mathbf{k}$. The quantity $\xi_{\mathbf{k}}$ is the
normal state energy dispersion. We adopt a six-parameter fit to the
band structure used previously for optimally doped Bi-2212
systems,~\cite{Norman95} having the form
\begin{eqnarray}
\xi_{\mathbf{k}}&=&-2t_{1} (\cos k_x + \cos k_y) -4t_{2} \cos k_{x}
\cos k_y \nonumber \\
&& -2t_{3} (\cos 2k_x + \cos 2k_y) \nonumber \\
&&-4t_{4} (\cos 2k_x \cos k_y + \cos k_x \cos 2k_y) \nonumber \\
&& -4 t_{5} \cos 2k_x \cos 2k_y - \mu \;,
\end{eqnarray}
where $t_1=1$, $t_{2}=-0.2749$, $t_{3}=0.0872$, $t_4=0.0938$,
$t_5=-0.0857$, and $\mu=-0.8772$. Unless specified explicitly, the
energy is measured in units of $t_1$ hereafter.

As in previous works,~\cite{Eschrig00,Zhu04,Sandvik04,Devereaux04}
we assume that an effective $d$-wave pairing interaction has
pre-existed from certain many-body effects. Therefore, even before
the electronic coupling to the bosonic excitations, the $d$-wave
superconducting order has already been established and the
corresponding order parameter is given by:
\begin{equation}
\Delta_{\mathbf{k}}=\frac{\Delta_{0}}{2}(\cos k_x -\cos k_y)\;.
\end{equation}
The additional renormalization comes from the additional
electron-bosonic mode interaction.

By introducing a two-component spinor operator, one defines a matrix
Green's function in the Nambu space. The bare Green's function  in
the assumption of a real $d$-wave pair potential reads
\begin{equation}
\hat{\mathcal{G}}_{0}^{-1}(\mathbf{k};i\omega_{n})=\left(
\begin{array}{cc}
i\omega_{n}-\xi_{\mathbf{k}} & -\Delta_{\mathbf{k}} \\
-\Delta_{\mathbf{k}} & i\omega_{n}+\xi_{\mathbf{k}}
\end{array}
\right)\;.
\end{equation}
Here $\omega_{n}=(2n+1)\pi T$ is the fermionic Matsubra frequency.

\subsection{Coupling to the collective modes}
\label{SUBSEC:Collective} As mentioned before, there are many
bosonic collective modes existing in cuprates. Here we focus on the
out-of-plane out-of-phase buckling ($B_{1g}$) and the in-plane
half-breathing phonon modes associated with the motion of the O
ions, and the $(\pi,\pi)$ spin resonance mode.

We model the electronic coupling to the phonon modes by
\begin{equation}
\mathcal{H}_{el-ph}=\frac{1}{\sqrt{N_{L}}}\sum_{\mathbf{k},\mathbf{q}
\atop \sigma} g_{\nu}(\mathbf{k},\mathbf{q})
c_{\mathbf{k}+\mathbf{q},\sigma}^{\dagger}
c_{\mathbf{k}\sigma}A_{\nu,\mathbf{q}}\;,
\end{equation}
where $N_{L}$ is the number of lattice sites,
$A_{\nu,\mathbf{q}}=b_{\nu\mathbf{q}} +
b_{\nu,-\mathbf{q}}^{\dagger}$ with $b_{\nu\mathbf{q}}^{\dagger}$
($b_{\nu\mathbf{q}}$) creating (annihilating) one phonon of type
$\nu$ (representing $B_{1g}$ or breathing mode) and wavevector
$\mathbf{q}$. We consider two types of the coupling matrix elements.
The first type are both $\mathbf{q}$ and $\mathbf{k}$ dependent:
\widetext
\begin{eqnarray}
g_{B_{1g}}(\mathbf{k},\mathbf{q})&=&\frac{g_{0}}{\sqrt{M(\mathbf{q})}}
\{\phi_{x}(\mathbf{k})\phi_{x}(\mathbf{k}+\mathbf{q}) \cos(q_y/2) -
\phi_{y}(\mathbf{k})
\phi_{y}(\mathbf{k}+\mathbf{q})\cos(q_x/2)\}\;,\\
g_{br}(\mathbf{k},\mathbf{q})
&=& g_{0} \sum_{\alpha=x,y}
\{\phi_{b}(\mathbf{k}+\mathbf{q})\phi_{\alpha}(\mathbf{k})\cos[(k_{\alpha}+q_{\alpha})/2]
-\phi_{b}(\mathbf{k})\phi_{\alpha}(\mathbf{k}+\mathbf{q})\cos(k_{\alpha}/2)\}\;,
\end{eqnarray}
\endwidetext
\noindent where $M(\mathbf{q})=[\cos^{2}(q_x/2) +
\cos^{2}(q_y/2)]/2$, and
\begin{eqnarray}
\phi_{x}&=&\frac{i}{\mathcal{N}_{\mathbf{k}}} \biggl{[}
\xi_{\mathbf{k}} t_{x,\mathbf{k}} -
t_{xy,\mathbf{k}}t_{y,\mathbf{k}}\biggr{]}\;,\\
\phi_{y}&=&\frac{i}{\mathcal{N}_{\mathbf{k}}} \biggl{[}
\xi_{\mathbf{k}} t_{y,\mathbf{k}} -
t_{xy,\mathbf{k}}t_{x,\mathbf{k}}\biggr{]}\;,\\
\phi_{b}&=&\frac{1}{\mathcal{N}_{\mathbf{k}}} \biggl{[}
\xi_{\mathbf{k}}^{2} - t_{xy,\mathbf{k}}^{2}\biggr{]}\;,
\end{eqnarray}
with  $\mathcal{N}_{\mathbf{k}}=\{ [\xi_{\mathbf{k}}^{2} -
t_{xy,\mathbf{k}}^{2}]^{2} + [\xi_{\mathbf{k}} t_{x,\mathbf{k}} -
t_{xy,\mathbf{k}}t_{y,\mathbf{k}}]^{2} + [\xi_{\mathbf{k}}
t_{y,\mathbf{k}} - t_{xy,\mathbf{k}}t_{x,\mathbf{k}}]^{2}\}^{1/2}$,
$t_{\alpha,\mathbf{k}}=-2t_{1}\sin(k_{\alpha}/2)$ and
$t_{xy,\mathbf{k}}=-4t_{2} \sin(k_x/2)\sin(k_y/2)$. The $\mathbf{k}$
dependence has been argued to be crucial in the interpretation of
ARPES data.~\cite{Cuk04,Devereaux04} The second type has only
$\mathbf{q}$
dependence:~\cite{Song95,Scalapino95,Nazarenko96,Dahm96,Bulut96,Sandvik04}
\begin{equation}
\vert g_{B_{1g}}(\mathbf{k},\mathbf{q})\vert^{2}=\vert g_{0}
\vert^{2} \biggl{[} \cos^{2}\biggl{(}\frac{q_x}{2} \biggr{)} +
\cos^{2} \biggl{(} \frac{q_y}{2}\biggr{)} \biggr{]}\;,
\end{equation}
\begin{equation}
\vert g_{br}(\mathbf{k},\mathbf{q})\vert^{2}=\vert g_{0} \vert^{2}
\biggl{[} \sin^{2}\biggl{(}\frac{q_x}{2} \biggr{)} + \sin^{2}
\biggl{(} \frac{q_y}{2}\biggr{)} \biggr{]}\;.
\end{equation}
This type of $\mathbf{q}$ has been the focus before in the context
of a $d_{x^2-y^2}$-wave pairing mechanism. For convenience of
notation, we refer to the phonon modes with the first type of
coupling as $B_{1g}$-I and $br$-I ones, while those with the second
type of coupling as $B_{1g}$-II and $br$-II ones.


The electronic coupling to the $(\pi,\pi)$ spin resonance mode is
modeled as:
\begin{equation}
\mathcal{H}_{el-sp}=\frac{g_0}{2N_{L}}\sum_{\mathbf{k},\mathbf{q}
\atop \sigma,\sigma^{\prime} }
c_{\mathbf{k}+\mathbf{q},\sigma}^{\dagger}(\mathbf{S}_{\mathbf{q}}\cdot
\bm{\sigma}_{\sigma\sigma^{\prime}})
c_{\mathbf{k},\sigma^{\prime}}\;,
\end{equation}
where $\mathbf{S}$ is the spin operator for the $(\pi,\pi)$ mode,
$\bm{\sigma}$ is the Pauli matrix vector.

We calculate the electronic self-energy due to the electron-bosonic
excitation coupling up to the second order in the coupling matrix
elements. For the electron-phonon coupling, the self-energy is given
as:
\begin{eqnarray}
\hat{\Sigma}(\mathbf{k};i\omega_{n})&=&-\frac{T}{N_{L}}\sum_{\mathbf{q},\nu}
\sum_{\Omega_{m}} g_{\nu}(\mathbf{k}-\mathbf{q},\mathbf{q})
g_{\nu}(\mathbf{k},-\mathbf{q}) \nonumber \\
&& \times  \mathcal{D}_{\nu}(\mathbf{q};i\Omega_{m}) \hat{\tau}_{3}
\hat{\mathcal{G}}_{0}(\mathbf{k}-\mathbf{q};i\omega_{n}-i\Omega_{m})\hat{\tau}_{3}\;,\nonumber
\\
\end{eqnarray}
where $\Omega_{m}=2m\pi T$ is the bosonic Matsubra frequency,
$\hat{\tau}_{3}$ is the third component of the Pauli matrix in the
Nambu space, the quantity
$\mathcal{D}_{\nu}(\mathbf{q};i\Omega_{m})$ is the Fourier transform
of the phonon Green's function
$\mathcal{D}_{\nu}(\mathbf{q};\tau)=-\langle T_{\tau}
[A_{\nu,\mathbf{q}}(\tau)A_{\nu,-\mathbf{q}}(0)]\rangle$ and is
taken as
\begin{equation}
\mathcal{D}_{\nu}(\mathbf{q};i\Omega_{m})=\frac{1}{2}
\left[\frac{1}{i\Omega_{m}-\Omega_{\nu}}-\frac{1}
{i\Omega_{m}+\Omega_{\nu}}\right]\;,
\end{equation}
with $\Omega_{\nu}$ being the resonance frequency of the phonon
modes.

For the electron-$(\pi,\pi)$-resonance  spin-fluctuation coupling,
the self-energy is given as:~\cite{Zhu04}
\begin{equation}
\hat{\Sigma}(\mathbf{k};i\omega_{n})=-\frac{3g_0^{2}T}{4N_{L}}
\sum_{\mathbf{q}} \sum_{\Omega_{m}} \chi(\mathbf{q};i\Omega_{m})
\hat{\mathcal{G}}_{0}(\mathbf{k}-\mathbf{q};i\omega_{n}-i\Omega_{m})
\;,
\end{equation}
where $\chi(\mathbf{q};i\Omega_{m})$ is the Fourier transform of the
spin-spin correlation function, $\chi(\mathbf{q};\tau)=-\langle
T_{\tau} [S_{\mathbf{q}}^{z}(\tau)S_{-\mathbf{q}}^{z}(0)]\rangle$
dynamical spin susceptibility. We treat the susceptibility in a
phenomenological form:~\cite{Eschrig00}
\begin{equation}
\chi(\mathbf{q};i\Omega_{m})=-\frac{f(\mathbf{q})}{2}
\left[\frac{1}{i\Omega_{m}-\Omega_{0}}-\frac{1}
{i\Omega_{m}+\Omega_{0}}\right]\;. \label{EQ:Suscep}
\end{equation}
Here the spin resonance mode energy is also denoted by $\Omega_0$.
The quantity $f(\mathbf{q})$ describes the momentum dependence of
the mode and is assumed to be enhanced at the $\mathbf{Q}=(\pi,\pi)$
point. Using the correlation length $\xi_{sf}$ (chosen to be 2
here), it can be written as
\begin{equation}
f(\mathbf{q})=\frac{1}{1+4\xi_{sf}^{2}[\cos^{2}\frac{q_{x}}{2} +
\cos^{2}\frac{q_{y}}{2}]}\;.
\end{equation}
The form of the dynamic susceptibility as given by
Eq.~(\ref{EQ:Suscep}) is especially suitable for the optimally doped
YBa$_2$Cu$_3$O$_{6+y}$ (YBCO) compounds in the superconducting
phase, where the observed neutron resonance peak is almost
resolution-limited in energy and fairly sharp in wavevector. The
resonance peak in BSCCO is broadened in both energy and wavevector.
In addition, given that the peak in BSCCO is still quite sharp in
energy, we expect that the energy broadening of the resonance mode
is not important for the present study. We have also neglected the
incommensurate peaks seen in the inelastic neutron scattering
experiments in YBCO (the part that disperses ``downward'' away from
the resonance peak),~\cite{Dai98,Arai99,Fong00,Brinckmann99,Kao00}
since their spectral weight is significantly smaller than that of
the resonance mode.

The dressed electron Green's function $\mathcal{G}$, due to the
renormalization effect of bosonic excitations, is given by:
\begin{equation}
\hat{\mathcal{G}}^{-1}(\mathbf{k};i\omega_{n})
=\hat{\mathcal{G}}_{0}^{-1}(\mathbf{k};i\omega_{n})
-\hat{\Sigma}(\mathbf{k};i\omega_{n})\;.
\end{equation}
To study the momentum transfer between the bosonic excitation
renormalized electronic states, additional impurities or defects are
required to scatter the electrons. The scattering from impurities is
described by,
\begin{equation}
H_{imp}=\sum_{l\sigma} U_{l} c_{l\sigma}^{\dagger} c_{l\sigma} \;,
\end{equation}
where $U_{l}$ is the strength of the zero-ranged impurity scattering
potential at the $l$-th site. For simplicity, we consider only the
case of nonmagnetic scattering and assume the scattering potential
from all these impurities are identical, i.e.,  $U_{l}=U_0$. The
full Green's function satisfies the following equation of motion:
\begin{equation}
\hat{G}(i,j;i\omega_{n}) = \hat{\mathcal{G}}(i,j;i\omega_{n}) +
\sum_{l} U_{l} \hat{\mathcal{G}}(i,l;i\omega_{n})\hat{\tau}_{3}
\hat{G}(l,j;i\omega_{n})\;.
\end{equation}
Since we are most interested in effects of the electron collective
mode coupling, it is desirable to keep as small as possible the
disturbance (i.e., quantum interference, formation of virtual
resonance etc.) arising from the quasiparticle scattering off the
impurities themselves. This suggests to consider a dilute
concentration of weak impurities. In this limit, the Born
approximation is applicable, and
  one arrives at:
\begin{equation}
\hat{G}(i,j;i\omega_{n}) = \hat{\mathcal{G}}(i,j;i\omega_{n}) +
U_{0}\sum_{l} \hat{\mathcal{G}}(i,l;i\omega_{n})\hat{\tau}_{3}
\hat{\mathcal{G}}(l,j;i\omega_{n})\;.
\end{equation}

Due to the impurity scattering, the correction to the local density
of states (LDOS)  at the $i$-th site, summed over two spin
components, is
\begin{equation}
\delta\rho(\mathbf{r}_{i},E)=-\frac{2U_0}{\pi}\sum_{l}
\mbox{Im}[\hat{\mathcal{G}}(i,l;E+i\gamma)\hat{\tau}_{3}
\hat{\mathcal{G}}(l,i;E+i\gamma)]_{11} \;. \label{ldos_correction}
\end{equation}


\subsection{Fourier Transform}
The local density of states is proportional to the local
differential tunneling conductance (i.e., $dI/dV$). To look into the
renormalization effect of collective bosonic excitations in the STM,
the energy derivative of the LDOS, corresponding to the derivative
of the local differential tunneling conductance (i.e.,
$d^{2}I/dV^{2}$), is more favorable to enhance the signal. For a
fixed value of energy, one first gets a set of $\delta
\rho^{\prime}(i,E)$ (the prime $\prime$ means the energy derivative)
in real space, and then performs the Fourier transform:
\begin{equation}
\delta \rho^{\prime}(\mathbf{q},E) = \sum_{i} \delta
\rho^{\prime}(\mathbf{r}_i,E) e^{-i\mathbf{q}\cdot
\mathbf{r}_{i}}\;,
\end{equation}
to obtain a map of the Fourier spectrum in $\mathbf{q}$ space,
\begin{equation}
P(\mathbf{q},E)=\vert \delta \rho^{\prime}(\mathbf{q},E) \vert\;.
\end{equation}

One can also prove that the relation between
$\delta\rho(\mathbf{q},E)$ and that due to a single impurity
$\delta\rho_{single}(\mathbf{q},E)$:
\begin{equation}
\delta\rho^{\prime}(\mathbf{q},E)=F(\mathbf{q})\delta\rho_{single}^{\prime}(\mathbf{q},E)\;,
\end{equation}
where $F(\mathbf{q})$ is the form factor for the spatial
distribution of weak impurities [$F(\mathbf{q})=1$ for the case of a
single impurity].


\section{Results and discussions}
\label{SEC:Results}

For the numerical calculation, we take the superconducting energy
gap $\Delta_0=0.1$, the frequency of all collective modes
$\Omega_{0}=0.15$. A quasiparticle lifetime broadening of
$\gamma=0.005$ is used. A weak impurity scattering strength
$U_0=0.1$. We take a large system size $N_{L}=1024\times 1024$ to
achieve the high-momentum and energy resolution. The Fourier
spectrum $P(\mathbf{q},E)$ is then constructed from $\delta
\rho^{\prime}(\mathbf{r}_i;E)$ within a given window of size
$61\times 61$ centering the single impurity. We choose the coupling
strength for all three types of collective modes in such a way that
at the Fermi energy $E=0$, the frequency renormalization factor $Z$
appearing in the self energy
\begin{equation}
\hat{\Sigma}(\mathbf{k};i\omega_{n})
=i\omega_{n}[1-Z(\mathbf{k};i\omega_{n})]\hat{\tau}_0 +
\chi(\mathbf{k};i\omega_{n})\hat{\tau}_{3} +
\Phi(\mathbf{k};i\omega_{n})\hat{\tau}_{1}\;,
\end{equation}
has the same real-part value for the $B_{1g}$-I, $B_{1g}$-II,
$br$-II phonon modes, and $(\pi,\pi)$-resonance spin fluctuation
modes at the $M$ point, while for the $br$-I phonon mode at the
wave-vector $(\pi/2,\pi/2)$. The calibrated value of the coupling
strength $g_0$ for all these collective modes is summarized in
Table~\ref{TAB:g0}.
\begin{table}
  \begin{tabular}{ccccc}
     \hline \hline
       $B_{1g}$-I mode & $br$-I mode & $B_{1g}$-II mode & $br$-II mode &
$(\pi,\pi)$ mode\\
     \hline
       1.05  & 1.48 & 0.75 & 0.75 & 2.30 \\
     \hline \hline
   \end{tabular}
\caption{The calibrated value of the coupling strength $g_0$ for
different type of collective modes.}
\label{TAB:g0}
\end{table}

\begin{figure}[th]
\centerline{\psfig{file=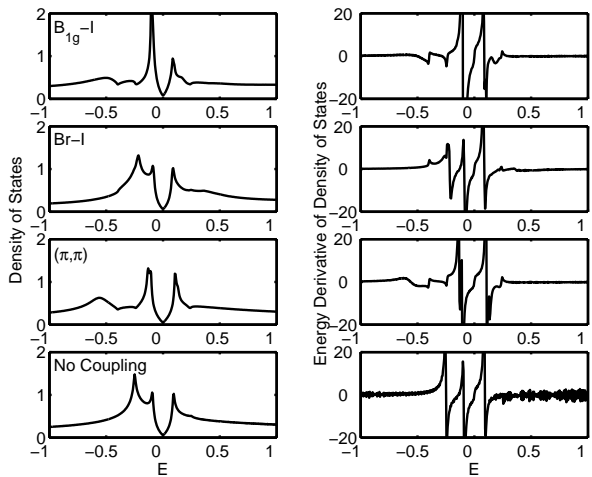,width=8cm}} \caption{Density of
states (left column) and its energy derivative (right column) as a
function of energy for a clean $d$-wave superconductor with the
electronic coupling to $B_{1g}$-I, br-I, and $(\pi,\pi)$ spin
resonance modes. The case of no mode coupling is also shown for
comparison. } \label{FIG:DOS}
\end{figure}

We present in Fig.~\ref{FIG:DOS} the results of the DOS and its
energy derivative as a function of energy for a clean (i.e.,
$U_{0}=0$) $d$-wave superconductor with the electronic coupling to
the $B_{1g}$-I, br-I, $(\pi,\pi)$ spin resonance modes. For
comparison, the DOS for the case of no mode coupling is also shown.
When there is no electron-mode coupling, there is a van Hove
singularity peak appearing outside the superconducting gap edge.
When the electrons are coupled to the $B_{1g}$ and $(\pi,\pi)$-spin
resonance modes, the van Hove singularity peak is strongly
suppressed. Instead, one sees a dip structure following the coherent
peak at the gap edge. The distance between this dip and the coherent
peak defines the resonance energy $\Omega_0$. However, there is very
little suppression when the electrons are coupled to the $br$-I
mode. The planar tunneling experiments indeed observed the peak-dip
structure rather than the peak-peak (van Hove singularity)
structure.
For the band-structure parametrization we have adopted,
this implies that
the electronic coupling to the collective modes must exist,
and the $B_{1g}$ and $(\pi,\pi)$
spin resonance mode are the most promising candidates.
Unfortunately, as shown in Fig.~\ref{FIG:DOS}, the dip structure due
to the coupling to the $B_{1g}$ and $(\pi,\pi)$ spin resonance mode
is almost identical. It would be very challenging to distinguish
between these two modes in the planar tunneling experiments, which
is measuring the momentum averaged DOS. Therefore, we propose to
look at the momentum transfer structure between the band
renormalized states, which can be measured by the FT-STM. To achieve
this goal, we need to have a signal strong enough to be detectable
in STM experiments. The derivative of the DOS, $\rho^{\prime}(E)$,
would serve the purpose. As shown in the right column of
Fig.~\ref{FIG:DOS}, when the electrons are coupled to the $B_{1g}$
and spin resonance modes, there is a strong peak structure at
$E=-(\Delta_0 + \Omega_0)$ exhibiting in the $\rho^{\prime}(E)$
spectrum, which has a one-to-one correspondence to the dip structure
in $\rho(E)$ itself.

\begin{figure}[th]
\centerline{\psfig{file=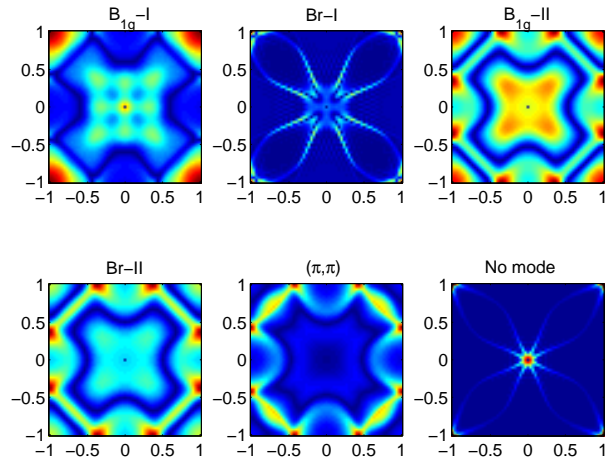,width=8cm}} \caption{The Fourier
spectral weight of the energy derivative of the LDOS at
$E=-(\Delta_0 + \Omega_0)$ for a $d$-wave superconductor with the
electronic coupling to the $B_{1g}$, br-I, $B_{1g}$-II, br-II, spin
resonace modes. For comparison, the quantity is also shown for the
case of no mode coupling. } \label{FIG:FOURIER-1}
\end{figure}

In Fig.~\ref{FIG:FOURIER-1}, we present the results of the Fourier
spectrum of the derivative of the LDOS,
$\rho^{\prime}(\mathbf{q},E)$, at the energy $E=-(\Delta_0 +
\Omega_0)$ for a $d$-wave superconductor with the electronic
coupling to the collective modes. For comparison, the same quantity
is also shown (last panel) for the case of no mode coupling. Note
that the case without the mode coupling, the energy $\Omega_0$ has
no special meaning in the context of the electronic properties, and
the energy $E=-(\Delta_0 + \Omega_0)$ is chosen merely for
comparison to the case of mode coupling. The main results are: For
all cases, there are strong intensity at the large momentum transfer
near $(\pi,\pi)$. For the cases of the 
$br$-I coupling and no mode
coupling,~\cite{footnote-no-mode-coupling} similar feature at very
small momentum transfer is obtained, relating to the fact that the
DOS spectrum in two cases (see Fig.~\ref{FIG:DOS}) is similar to
each other. For the cases of the coupling to the $B_{1g}$-II and
$br$-II modes, there are intensity peaks at a finite momentum
transfer along the diagonals. For the case of the coupling to the
spin resonance mode, no strong feature is obtained at the
intermediate value of momentum transfer, which is consistent with
our previous calculation for this specific case.~\cite{Zhu04} For
the case of the coupling to the $B_{1g}$-I phonon mode, there exists
not only the intensity peaks with a momentum transfer along the
diagonal but also the ones with a momentum transfer along the bond
directions of a square lattice.

\begin{figure}[th]
\centerline{\psfig{file=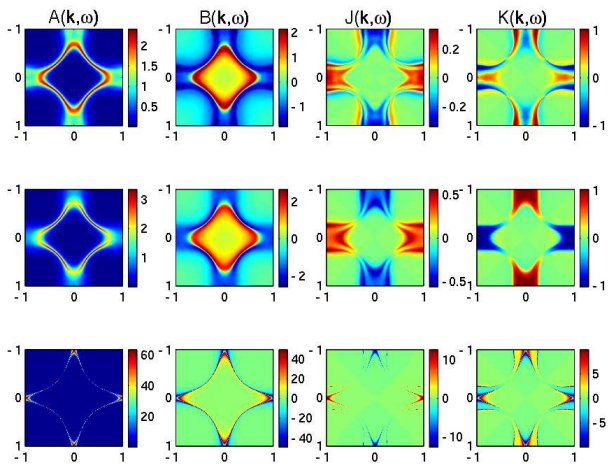,width=8cm}} \caption{The imaginary
and real parts of  the single particle and anomalous Green's
function, $A(\mathbf{k};E)$, $B(\mathbf{k};E)$, $J(\mathbf{k};E)$,
and $K(\mathbf{k};E)$ [as defined by
Eqs.~(\ref{EQ:A})-(\ref{EQ:K})], at the energy $E=-(\Delta_0 +
\Omega_0)$ for the electronic coupling to the $B_{1g}$ phonon (first
row), spin resonance mode (second row), and the case of no mode
coupling. } \label{FIG:ARPES}
\end{figure}

The different Fourier spectrum patterns come from the detailed
renormalization of electronic structure by the coupling to these
modes. To better understand these patterns, we turn to a detailed
analytical form of the Fourier transform.
By putting aside the external form factor
associated with a specific configuration of weak disorder, the Fourier
spectrum is determined uniquely by the electronic single particle
Green's function and is found to be:
\begin{eqnarray}
&\delta\rho(\mathbf{q};E)=\frac{u_0}{N_{L}} \sum_{\mathbf{k}} &
\nonumber \\
& \{ [A(\mathbf{k};E)B(\mathbf{k}+\mathbf{q};E) +
B(\mathbf{k};E)A(\mathbf{k}+\mathbf{q};E)] & \nonumber \\
& -[J(\mathbf{k};E)K(\mathbf{k}+\mathbf{q};E) +
K(\mathbf{k};E)J(\mathbf{k}+\mathbf{q};E)]\}\;,& \label{EQ:FT}
\end{eqnarray}
where
\begin{eqnarray}
A(\mathbf{k};E)&=&-\frac{2}{\pi}
\mbox{Im}[\mathcal{G}_{11}(\mathbf{k};E+i\gamma)]\;,
\label{EQ:A}\\
B(\mathbf{k};E)&=&\mbox{Re}[\mathcal{G}_{11}(\mathbf{k};E+i\gamma)]\;,
\label{EQ:B} \\
J(\mathbf{k};E)&=&-\frac{2}{\pi}
\mbox{Im}[\mathcal{G}_{12}(\mathbf{k};E+i\gamma)]\;,
\label{EQ:J} \\
K(\mathbf{k};E)&=&
\mbox{Re}[\mathcal{G}_{12}(\mathbf{k};E+i\gamma)]\;. \label{EQ:K}
\end{eqnarray}
Here as shown in Eqs.~(\ref{EQ:FT}) through~(\ref{EQ:K}), the
Fourier spectrum is determined by the convolution of the
imaginary~\cite{footnote-arpes} and the real parts of the
single-particle $(\mathcal{G}_{11})$ and anomalous
$(\mathcal{G}_{12})$ Green's function in the superconducting state.
The stronger intensity in $\delta\rho(\mathbf{q};E)$ (also
$\delta\rho^{\prime}(\mathbf{q};E)$) will be contributed from the
wave vector $\mathbf{q}$, which connects the simultaneously largest
intensity in $A(\mathbf{k};E)$ and $B(\mathbf{k};E)$ maps, and in
$J(\mathbf{k};E)$ and $K(\mathbf{k};E)$ maps. To be illustrative, we
present in Fig.~\ref{FIG:ARPES} those maps for the electronic
coupling to the $B_{1g}$-I and spin resonance modes, and also for
the case without the mode coupling at $E=-(\Delta_0 + \Omega_{0})$.
Notice that the joint intensity of
$J(\mathbf{k};E)K(\mathbf{k}^{\prime};E)$ are smaller by an overall
factor of 10 than that of $A(\mathbf{k};E)B(\mathbf{k}^{\prime};E)$
and the Fourier spectrum is mostly determined by the latter product.
For the case of no mode coupling, the strongest weight in these
quantities are located at the $M$ points of the first Brillouin
zone, which leads to the strongest intensity of the Fourier spectrum
with the momentum transfer $\mathbf{q}=(\pi,\pi)$ and the zero
momentum transfer $\mathbf{q}=0$. If the electrons are coupled to
the $B_{1g}$-I phonon mode, besides the strongest intensity on the
closed ridges in $A(\mathbf{k};E)$ and $B(\mathbf{k};E)$ [red area],
there are also moderately strong intensity on the two split beams
around the $M$ points in these two maps [bright green in the former
and dark blue in the latter]. The intensity
$A(\mathbf{k};E)B(\mathbf{k}^{\prime};E)$ connected by
$\mathbf{q}=\mathbf{k}^{\prime}-\mathbf{k}$ with $\mathbf{k}$ and
$\mathbf{k}^{\prime}$ located at the ends of these beams becomes
stronger. These wave vectors are just those in the $B_{1g}$-I panel
of Fig.~\ref{FIG:FOURIER-1} at which the Fourier spectrum exhibits
peaks [green spots]. However, if the electrons are coupled to the
spin resonance mode, no such beams exist, which explains the lack of
peaks in Fourier spectrum [see the $(\pi,\pi)$ panel in
Fig.~\ref{FIG:FOURIER-1}].

\begin{figure}[th]
\centerline{\psfig{file=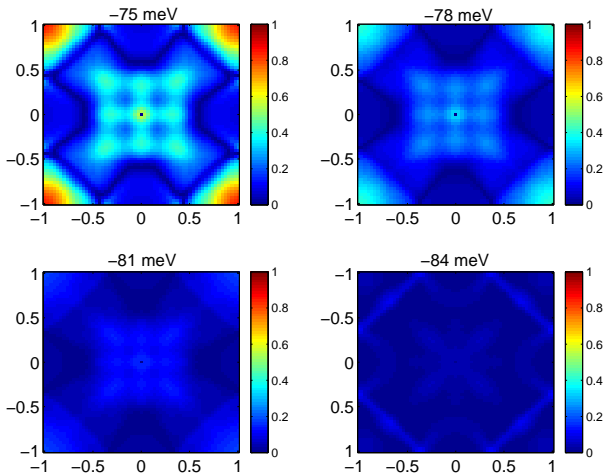,width=8cm}} \caption{The Fourier
spectrum of the derivative of the LDOS is shown at the various
values of the energy for the case of the electronic coupling to the
$B_{1g}$ phonon mode. Here the energy has been measured by scaling
$\Delta_0=30\;\mbox{meV}$.} \label{FIG:FOURIER-2}
\end{figure}

Experimentally, only the peaks in the Fourier spectrum with the
momentum transfer along the bond direction has been observed so
far.~\cite{JLee05} The scenario that the electrons are coupled to
the $B_{1g}$ phonon mode and the coupling matrix is highly
anisotropic (i.e., $B_{1g}$-I mode) bears the closest resemblance to the
experimental observation, though not in a full agreement. In
Fig.~\ref{FIG:FOURIER-2}, we present the energy evolution of the
spectrum pattern for the electronic coupling to the $B_{1g}$-I mode.
It shows that the characteristic momentum transfer wave vector
decreases slightly with the increased energy. However, the intensity
at these wave vectors decreases rapidly when the energy moves away
from the action point $-(\Delta_0 +\Omega_0)$. This is also not
inconsistent with the experiment.

\section{Concluding remarks}
\label{SEC:Conclusion}

There is considerable evidence that numerous ARPES and tunneling data
can be interpreted in terms of the electronic coupling to bosonic modes
with energy about 40 meV. Possible candidates for this mode are the
$(\pi,\pi)$ spin resonance mode and various phonon modes. The planar
junction tunneling has provided an accurate measurement of the
energy scale of this mode. However, since the structure in the
tunneling spectra due to the electronic coupling to $B_{1g}$ phonon
mode or to the $(\pi,\pi)$ spin resonance mode does not have
much difference, it is very difficult to determine the nature
of the mode.

In this paper, we have analyzed the Fourier spectra of the energy
derivative LDOS to investigate the momentum transfer structure
arising from the electronic coupling to these different modes.
This quantity nicely complements ARPES for the understanding
of the electronic responses to these modes. It can now be obtained from
the FT-STM experiments with
elevated spatial resolution.
In general, we found that the detailed momentum dependence
of the coupling matrix element strongly influences the electronic
properties. In
particular, we have shown that if the $B_{1g}$ or breathing modes
are coupled to the electrons with only a $\mathbf{q}$ dependence,
the spectrum displays the peak structure along the diagonals with a
small momentum transfer while the $(\pi,\pi)$ mode coupling does not
produce much weight at the small momentum transfer region.
On the other hand, if the electrons are coupled to the $B_{1g}$
phonon mode with a matrix element that depends not only on $\mathbf{q}$
but also on $\mathbf{k}$,
the peak structure with a small momentum transfer
can also appear along the Cu-O bond directions of the CuO$_{\rm 2}$ plane.
Recent FT-STM experiments have indeed found a peak at a
bond-directed momentum transfer.

Our
calculations also show the structures at large momentum transfer for
all cases of the electron-collective mode coupling.
In other words, both types (electronic and phononic) of mode couplings
produce structures near $(\pi,\pi)$, but only a coupling to
the phonon modes yields additional structures at the
small momentum transfers. No peak structures near $(\pi,\pi)$
have been observed in the experiments.
The situation is somewhat similar to the elastic scattering case
(i.e., in the absence of the collective mode coupling),
where similar structures near $(\pi,\pi)$ also appear
in the theoretical spectra~\cite{Wang02,Zhang03} but are not
observed experimentally. It is likely that the lack of structures
at large momentum-transfer in the elastic and inelastic experiments
has a common origin. One possibility has to do with
strong inhomogeneities, which may give rise to a dominant forward
scattering and make only the structures at small momentum transfers
observable. This amounts to the following form factor for
the weak disorder configuration:
\begin{equation}
F({\mathbf{q}})=\frac{1}{1 + r_c [\sin^{2}(q_x/2) +
\sin^{2}(q_y/2)]}\;,
\end{equation}
where the parameter $r_c$ controls the range of the forward
scattering in the $\mathbf{q}$ space. The overall modulation
in the Fourier spectrum of the derivative LDOS
$\delta \rho^{\prime}({\bf q}, \omega) \propto F(\mathbf{q})$
will be confined to small momenta if $F(\mathbf{q})$ is.

Finally, several remarks are in order: (i) The electron-collective
mode coupling we have considered preserves the translational
symmetry. Fourier spectra were studied after introducing elastic
impurity scattering with weak scattering potential. Alternatively,
local vibrational mode scattering  will not only provide an
inelastic scattering channel but also will break the translational
symmetry at the beginning.~\cite{Balatsky03} To fully understand the
possible FT-STM experiments, it would be instructive to consider the
electronic coupling to the distributed local vibrational modes. We
leave this problem for separate analysis. (ii) There has also be
increased interest in the quasiparticle scattering from the $\tau_1$
impurities.~\cite{Nunner05} It would be very helpful to study the
the FT spectrum through the $\tau_1$ scatterers. We delay this
investigation to a future publication.

\acknowledgments

We thank D.-H. Lee, N. Nagaosa, M. R. Norman, D. J. Scalapino, and
Z. X. Shen for very useful discussions. This work was supported by
the US DOE (J.X.Z. and A.V.B.), the NSERC, the Office of Naval
Research under Grant No. N00014-05-1-0127, and the A. von Humboldt
Foundation (T.P.D.), the NSF under Grant No. DMR-0424125 and the
Robert A. Welch Foundation (Q.S.), the Office of Naval Research
under grant N00014-03-1-0674, the NSF under Grant No. DMR-9971502,
the NSF-ITR FDP-0205641, and the Army Research Office under Grant
No. DAAD19-02-1-0043 (K.M., J.L., and J.C.D).


\begin{thebibliography}{99}
\bibitem{footnote-no-mode-coupling} Here, with the experiment of
Ref.~\onlinecite{JLee05} in mind, we have done a Fourier transform
of the LDOS in the entire field of view. If impurities are
explicitly identified, as in the case of Zn-doped cuprates, a
filtered Fourier transform can also be adopted. As discussed in
Ref.~\onlinecite{Zhu04}, this procedure would consider the LDOS at
the sites other than those of two diagonal strips; in other words,
the Fourier transform would only incorporate sites from four
disconnected trangles. Within this filtered Fourier transform
procedure, the features near $(\pi,\pi)$ disappear in the absence of
any mode coupling, but persist for all the cases involving a mode
coupling.
\bibitem{footnote-arpes}
Notice that the spectral weight $A(\mathbf{k},\omega)$ as given
by Eq.~(\ref{EQ:A}) is the quantity that the ARPES experiments can
measure. The role of the $(\pi,\pi)$-resonance spin fluctuation and
the phonon modes on the superconducting state spectral function has
been studied previously. This function gives the information about
the single particle energy dispersion in the first Brillouin zone in
a homogeneous system.
\end{thebibliography}
\end{document}